\documentclass[sigconf, authorversion, nonacm]{acmart}

\usepackage{booktabs}

\setcitestyle{nosort,square} 

\usepackage{hyperref}
\usepackage{amsmath}
\usepackage{graphicx}
\usepackage{comment}
\usepackage{soul}
\usepackage{multirow,bigdelim}
\usepackage{tabularx}
\usepackage{xcolor}
\usepackage{array}
\usepackage{wrapfig}
\usepackage{rotating}
\usepackage{xfp}
\usepackage[position=top]{subfig}
\usepackage{floatflt}
\usepackage{csvsimple}
\usepackage{adjustbox}
\usepackage[percent]{overpic}
\usepackage{float}
\usepackage{mathtools}
 \usepackage{booktabs}
 \usepackage{balance}

\newcolumntype{C}[1]{>{\centering\let\newline\\\arraybackslash\hspace{0pt}}m{#1}}

\newcommand{\ourmethod}{Mesh Draping}
\newcommand{\ours}{Mesh Draping}
\newcommand{\src}{source}
\newcommand{\tgt}{target}
\newcommand{\msrc}{\mathcal{M}}
\newcommand{\omsrc}{\widehat{\mathcal{M}}}
\newcommand{\mtgt}{\mathcal{T}}
\newcommand{\mloss}{\mathcal{L}}

\newcommand{\quality}{$\mathbf{Q}_\text{transfer}$}
\newcommand{\mfunc}{\mathcal{F}}

\newcommand{\ra}[1]{\renewcommand{\arraystretch}{#1}}

\newcommand{\net}{FC network}

\newcommand{\elc}{ELC}  
\newcommand{\ism}{ISM}  
\newcommand{\ncage}{NC}  
\newcommand{\sflow}{SF}  
\newif\ifdraft
\draftfalse

\ifdraft
\newcommand{\dcc}[1]{{\color{red}[\textbf{DC:} #1]}}
\newcommand{\ahc}[1]{{\color{orange}[\textbf{AH:} #1]}}
\newcommand{\rgc}[1]{{\color{green}[\textbf{RG:} #1]}}
\newcommand{\opc}[1]{{\color{blue}[\textbf{OP} #1]}}

\newcommand{\ah}[1]{{\color{orange}#1}}

\newcommand{\op}[1]{{\color{blue}#1}}

\else
\newcommand{\dcc}[1]{}
\newcommand{\rgc}[1]{}
\newcommand{\ahc}[1]{}
\newcommand{\opc}[1]{}
\newcommand{\osc}[1]{}

\newcommand{\ah}[1]{{\color{black}#1}}
\newcommand{\op}[1]{{\color{black}#1}}

\fi

\acmJournal{FACMP}

\begin{document}
\title{Mesh Draping:  Parametrization-Free Neural Mesh Transfer}
\author{Amir Hertz}
\affiliation{\institution{Tel Aviv University}\country{ }}
\email{amirhertz@mail.tau.ac.il}
\author{Or Perel}
\affiliation{\institution{Tel Aviv University}\country{ }}
\email{orr.perel@gmail.com}

\author{Raja Giryes}
\affiliation{\institution{Tel Aviv University}\country{ }}
\email{raja@tauex.tau.ac.il}
\author{Olga Sorkine-Hornung}
\affiliation{\institution{ETH Zurich, Switzerland}\country{ }}
\email{sorkine@inf.ethz.ch}
\author{Daniel Cohen-Or}
\affiliation{\institution{Tel Aviv University\country{ }}}
\email{cohenor@gmail.com}
\renewcommand{\shortauthors}{Hertz, A. et al.}

\begin{abstract}
Despite recent advances in geometric modeling, 3D mesh modeling still involves a considerable amount of manual labor by experts. In this paper, we introduce \ourmethod{}: a neural method for transferring existing mesh structure from one shape to another. 
The method drapes the source mesh over the target geometry and at the same time seeks to preserve the carefully designed characteristics of the source mesh. 
At its core, our method deforms the source mesh using progressive positional encoding. We show that by leveraging gradually increasing frequencies to guide the neural optimization, we are able to achieve stable and high quality mesh transfer. Our approach is simple and requires little user guidance, compared to contemporary surface mapping techniques which rely on parametrization or careful manual tuning. Most importantly, \ourmethod{} is a parameterization-free method, and thus applicable to a variety of target shape representations, including point clouds, polygon soups, and non-manifold meshes.
We demonstrate that the transferred meshing remains faithful to the source mesh design characteristics, and at the same time fits the target geometry well.

\end{abstract}

\begin{teaserfigure}
\vspace{-1mm}
 \includegraphics[width=0.9\linewidth]{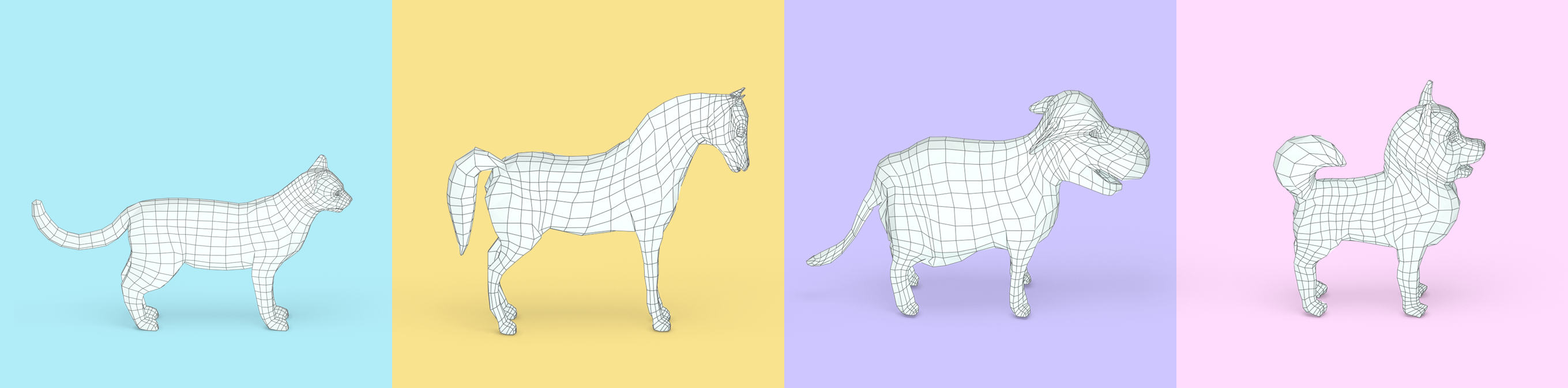}
 \centering
 \vspace{-1mm}
  \caption{\ourmethod{} in action: the source mesh of the cat model (left), is transferred with minimal distortion onto other target shapes.}
  \vspace{1mm}
\label{fig:teaser}
\end{teaserfigure}
\maketitle

\section{Introduction}

Polygonal meshes are de facto the most common discrete representation of surfaces in computer graphics. For geometric modeling purposes, their expressiveness allows to capture adequate approximations of many object surfaces. At the same time, their modular design has sparked a plethora of interactive editing tools. That, in turn, makes them appealing to 3D artists for various creative tasks, such as modeling, sculpting, and rigging.

The flexible nature of meshes allows artists to conceive shapes of varying densities by diligently employing local operations. Detailed mesh areas, and in particular curved parts and features relevant for articulation, are usually carefully crafted with higher polygonal count and certain connectivity structures. For example, a mesh of a humanoid may contain both a detailed, dense polygon fan of an eye, as well as a low-poly,  flat torso. 
Much of this delicate design involves meticulous repetitive work, oftentimes when jointly drafting multiple mesh models. Experts may resort to the usage of templates, but this is limited to certain use cases. A scanned point cloud, for instance, is not generated from such a template and therefore cannot directly inherit a desired mesh topology. 
To alleviate the hard labor of artists, a wealth of research efforts have been focusing on \emph{meshing} and \emph{remeshing} techniques, which, given a \tgt{} polygonal shape, generate an alternative, coherent and enhanced mesh topology. This is usually achieved by optimization methods with various objectives that aim to optimize the polygon distribution while avoiding degenerate faces~\cite{campen2017partitioning}.

While these methods are useful, they are inadequate at times, as they neither allow users full control over the resulting connectivity, nor directly support polygon soups and point cloud targets. As a consequence, they are missing the opportunity to re-use sophisticated meshes carefully crafted by experts.

In this paper, we introduce an alternative approach, namely, a method for transferring an existing, high quality reference mesh structure to a comparable \tgt{} shape. Unlike previous methods that optimize both the connectivity and the geometry of the target shape mesh, our method reuses the source mesh characteristics while deforming it to best fit the geometry of the target. Our algorithm relies on a neural optimization process at its core: we learn the weights of a lightweight network to deform a given source mesh to the target shape.  \ah{In Figure~\ref{fig:teaser}, we show a mesh transfer from a source
surface left to three different target shapes.}
We follow contemporary literature, and overcome the shortcomings of simple neural nets by employing varying positional encodings. More specifically, we gradually map the source mesh vertex positions though encoding functionals of increasing frequencies ~\cite{hertz2021sape} from low to high, before feeding them into the network. This progressive scheme introduces an \textit{inductive bias} of slowly changing global deformations. That allows us to maintain a stable optimization through the earlier, critical steps of the optimization. Towards the end of the optimization, it also allows us to fit delicate details, where higher frequencies pertaining to fine local features are mapped.

We propose \ourmethod{}, a parameterization-free, neural approach for mesh transfer between shapes. \ourmethod{} is an interactive tool, where the user merely guides the displacement procedure. That is, the source mesh is “draped” over the target geometry. Our optimizing technique yields high quality meshes, faithful to both the source design and the geometry of the target shape. Most importantly, it is not limited by the target shape representation, rendering our approach suitable for transfering mesh structures to an assortment of shape formats, such as point clouds, polygon soups, and nonmanifold meshes. We demonstrate results of transferring triangle and quad meshes on a variety of examples, ranging from 3D models generated by artists to synthetic objects comparable with other methods.
\section{Related Work}

\subsection{Remeshing}
The literature on \emph{meshing} and \emph{remeshing} is concerned with methods that directly generate new mesh structures of the desired shapes without assuming any prescribed connectivity. Oftentimes, a generated mesh of good quality ideally comprises \ah{of} well-shaped triangles or quadrilaterals that are aligned to principal directions.

Different remeshing methods allow varying degrees of control by the user.
Automated or semi-automated techniques may utilize high level inputs, such as density control and crease selection \cite{alliez2002interactive,lai2008incremenral,fang2018quadrangulation,jakob2015instant}.
Other methods allow further interactions with the user. \cite{bommes2009mixed} let users specify orientation and alignment constraints. \cite{campen2014dual} introduce dual strip weaving, a framework where the user can specify the main layout of the quad grid. \cite{marcias2015data} learn quadrangulation patterns, applying them by guided user sketches of the edge flows. \cite{ebke2016interactively} suggest a rapid feedback interface that enables more precise control by specifying constraints,  such as the placement of singularities and their connections, as well as smooth surface regions, among others. The above methods generate high quality meshes, however, they do not offer users the option of employing  existing, custom mesh structures. 

Another class of remeshing methods deals with compatible meshing. Such methods remesh multiple shapes to generate common connectivity among them \cite{yang2018volume,yang2020error}. These methods generally aim to produce meshes of regular connectivity and minimal polygon count that fit both shapes to facilitate applications like morphing or attribute mapping; however, they discard the original mesh connectivity, in contrast to the goal of our work.

For further reading about triangle and quadrilateral remeshing algorithms, we refer the curious reader to the comprehensive surveys by \cite{alliez2008recent} and \cite{campen2017partitioning}.

\subsection{Surface Mapping}

Our work shares similarities with methods that map attributes between surfaces.
In the following, we provide a brief summary of relevant surface mapping techniques; a more extensive discussion is available in the survey by \cite{tam2013registration}.

Broadly speaking, most works can be characterized into extrinsic approaches, which leverage surface deformation embedded within a 3D space, and intrinsic approaches, which achieve a complete bijective mapping between the two surfaces via parameterization to some intermediate domain.

In a purely extrinsic approach, the deformation process is preceded by global alignment defined by corresponding points on the source and target shapes  \cite{zhang2008deformation, li2008global}. Local transformations calculated per element depend on the global orientation of both shapes.
Early approaches, such as the renowned iterative closest point method (ICP)  \cite{chen1992icp,besl1992icp}, assume the scenario of rigid registration, that is, the transformation between the source and target shapes maintains pairwise point distances.
Follow up works remove this restriction and perform non-rigid registration \cite{sharf2006SnapPasteAI}, possibly in a setting of partially overlapping surfaces with outliers \cite{li2008global,huang2009nonrigid,bouaziz2013sparse}.

By contrast, in the intrinsic line of works, mapping is achieved by means of parameterization to an intermediate domain. The actual mapping is achieved using a composition of mappings from source shape to the intermediate domain and an inverse mapping from the intermediate domain to the target shape. Representative examples include
\cite{kraevoy2004cross, kim2011blended, aigerman2015seamless, aigerman2015orbifold, aigerman2016hyperbolic, baden2018mobius, schmidt2019distortion}. Many of these works aim for either continuous or bijective mappings, or both.
To handle non-isometric cases, \cite{mandad2017variance} suggest using an optimal transport plan without requiring an intermediate domain, geared towards as-conformal-as-possible mapping. A recent work by \cite{schmidt2020inter} presents a novel method for continuous bijective mapping (homeomorphism) of attributes between surfaces of the same genus. Their method is able to obtain low intrinsic distortion, as well as  generalize to arbitrary genus. \cite{deng2020better} present a neural approach to reconstruct a target shape via an atlas of stitched patch mappings. Unlike our method, these methods rely on surface parameterization and cannot be easily extended to domains beyond manifold meshes, such as point clouds.

\cite{ezuz2019elastic} propose a hybrid extrinsic approach that  builds upon an intrinsic initialization. Their design combines an optimization scheme of elastic thin-shell deformation of the source mesh with projection over the target. They can handle non-isometric shapes, but struggle on highly complex meshes. 

The problem of non-isometric mapping has also been studied from a spectral point of view. \cite{ovsjanikov2012functional,ovsjanikov2017computing} propose using functional maps to transfer functions between two meshes using eigendecomposition. Newer neural variants of this method also exist \cite{litany2017deepfunctional, ginzburg2020cyclicFM, donati2020deepgeometricfunctional}. These works create vertex-to-an-area mappings, rather than vertex-to-vertex mappings, which renders them less suitable for the purpose of mesh transfer.

The work of \cite{tierny2011inspired} is closest to ours in spirit, as it also allows for direct quad mesh transfer between source and target meshes. In their work, they  construct a corpus of source meshes, and use cross-parameterization with as-rigid-as-possible deformations to to automatically generate a mapping between the shapes. Their method assumes a parameterization of the target shape is available, as well as the existence of a homeomorphism. Their method requires a preprocessing phase for generating the aforementioned corpus, where boundary conditions are provided and segmentation masks are marked.
Their method then requires to  stitch the back the segments in a postprocess.

A common application for surface mapping methods is texture mapping, where the  emphasis is on minimizing global distortion of the mapped textures. In contrast, our work is parameterization-free, global, and differs by focusing on correlative face transfer between similar areas of the two shapes, denoted by users, as well as maintaining the original mesh integrity. While mesh transfer can be implemented via a surface mapping method, we show in Section  \ref{section:experiments} that such methods produce inferior results in the extrinsic case,
and cannot be used 
when a manifold mesh is not available and bijective parameterization cannot be directly achieved.

\subsection{Shape Deformation} 
%
Applying unconstrained deformations to mesh vertices introduces a risk of geometric artifacts, such as intersecting faces.
Our work shares common ground with shape deformation methods, and similarly to them, we cope with the challenge of maintaining the mesh coherency under geometric warping.
To that end, many of the shape deformation methods are concerned with the type of regularization imposed on the mesh transformations, or proper design of interactive editing tools offered to the user.
We highlight however, that the end goal of shape deformation works is somewhat different: they aim to preserve a single source shape under some set of user constraints. For an elaborate review of non-neural methods, we refer readers to \cite{DeformationTutorial:2009}, \cite{jacobson2013algorithmsai} and the references therein.

%
%

To tie up the discussion, we mention works most relevant to our problem domain. 
\cite{TemplateMeshFitting:2010} suggest deformation by template based fitting of 3D meshes, using a coarse to fine pipeline. 
\cite{jacobson2011bounded} present a unified interactive approach combining control handles, skeletons and cages to deform a given shape. Their method uses weighted linear blending for practical purposes of mesh editing and rigging.


More recently, \cite{yifan2020neural} demonstrated a learned variant of cage based deformation. Given a source and a target shape, their algorithm warps the source mesh to match the target. However, this work focuses on preserving local details within the source shape, and does not guarantee that the result intimately coincides with the target shape.
Another contemporary work by \cite{wang20193dn} performs neural warping of source to target shapes by concatenating their global representations and predicting per-vertex offsets. Their method relies on a compound loss that assumes symmetry, and is limited to the domain of the training set.
The ShapeFlow method \cite{jiang2020shapeflow} is a scheme for learning a deformation space. Given a target shape, the approach employs a nearest neighbor search for a pre-learned source candidate to guide the warp process.
Our method, in contrast, does not assume any learned shape representation, and can be applied directly to novel shapes from unseen domains.

\newcommand{\offsetxd}{81.5}
\begin{figure*}
\begin{overpic}[width=\textwidth,tics=10]{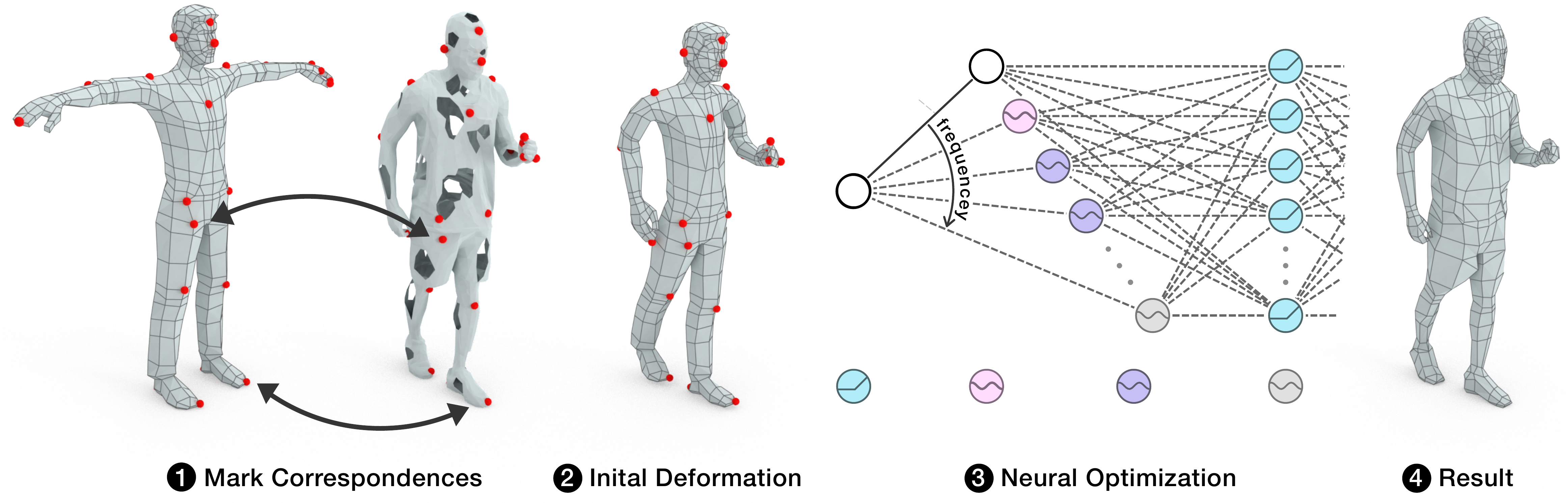}

\put(54,19.7){$p$}
\put(62.4,27.6){$p$}

\put(5.4,6){\small{Source}}
\put(23,6){\small{Target}}
\put(56,6.9){\small{ReLU}}
\put(64.4,6.9){\small{Periodic}}
\put(73.6,6.9){\small{In progress}}
\put(83.5,6.9){\small{Off}}

\end{overpic}
\caption{Method Overview. Our pipeline beings with a manual step of marking sparse correspondence points between the source and the target. Then, source mesh deformation takes place, to obtain an initial solution for the mesh transfer neural optimization. After that, the optimization commences, until it finally converges to the resulting transferred mesh.}
\label{fig:ppe_diagrm}

\end{figure*}
\begin{figure*}
\begin{overpic}[width=\textwidth,tics=10, trim=0 40 0 0, clip]{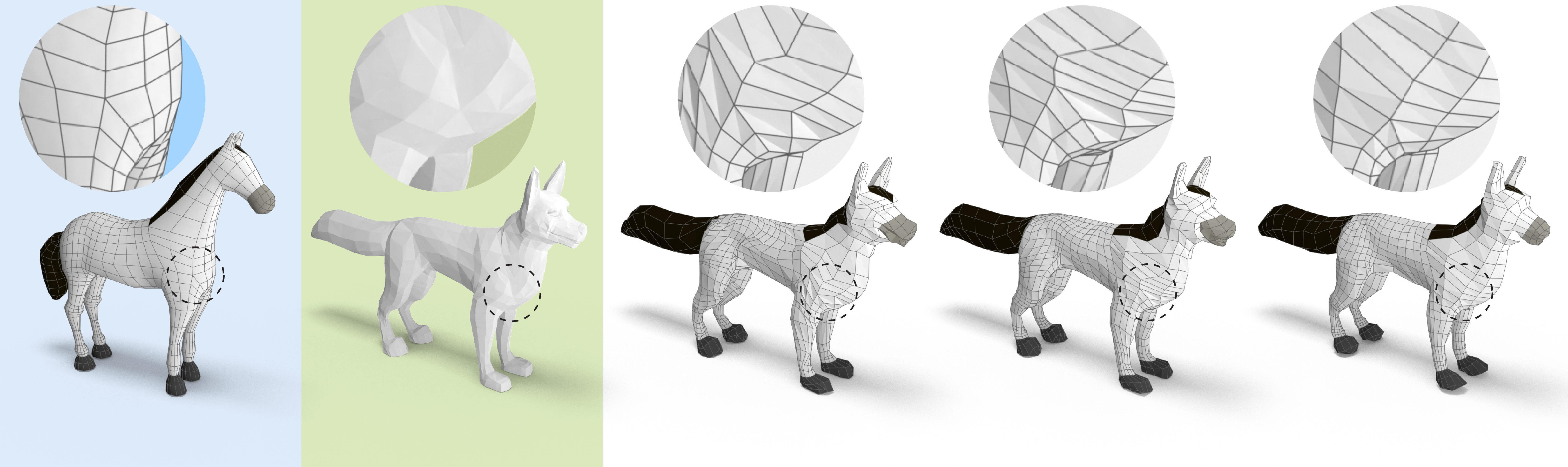}

\put(7,1){Source}
\put(26.5,1){{Target}}

\put(44.5,1){Area $D_{kl}$}
\put(67,1){Angles}
\put(83.5, 1){Angles + Area $D_{kl}$}
\end{overpic}
     \caption{\op{Ablation of the structural loss component: $\mloss_{s}\left(\omsrc | \msrc\right)$ (Eq.~\ref{eq:structural_loss}).
     Middle: $\mloss_{s}$ considers only the area term, $D_{kl}$, which leads to large distortions with respect to the source mesh (blue). Next to the right, optimizing while minimizing the angle error term achieves better results. Right: combining both losses better preserves the tessellation of the source mesh. 
     }}
     \label{fig:ablation_loss}
\end{figure*}
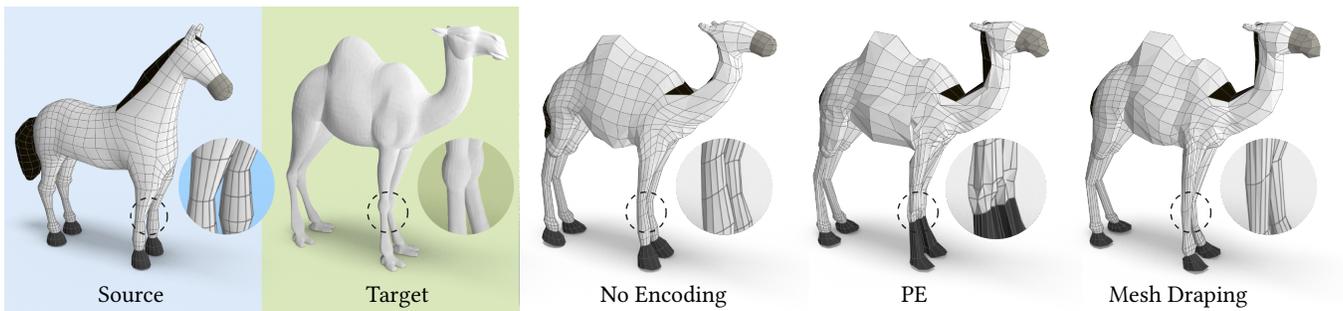
\begin{figure*}
\begin{overpic}[width=\textwidth,tics=10]{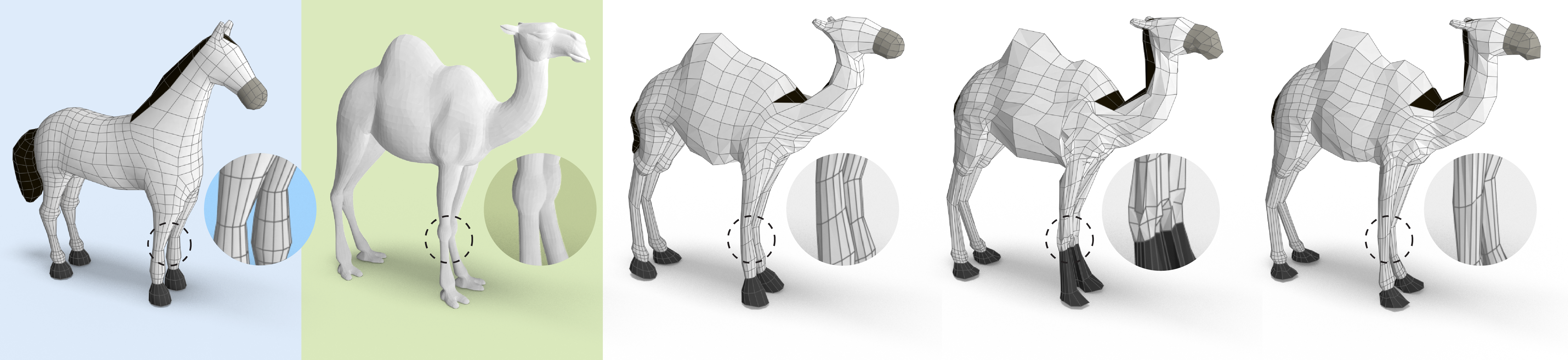}

\put(7,1){Source}
\put(27,1){{Target}}

\put(44.5,1){No Encoding}
\put(67,1){PE}
\put(82.5, 1){\ourmethod{}}
\end{overpic}
     \caption{Mesh transfer from source to target, comparing various neural optimization methods. No Encoding, which uses common MLPs with ReLU activations, fails to fit delicate target geometry details. Positional Encoding is able to fit fine details, but converges to a sub-optimal minima which introduces distortions. Progressive Positional Encoding endows a coarse-to-fine inductive bias, which contributes to stable optimization and and the ability to achieve high quality mesh transfer.}
     \label{fig:model_compare}
\end{figure*}

\section{Method Overview}
\label{sec:method}

Our framework begins with a preliminary step, where the user specifies a small number of correspondence points on the \src{} and \tgt{} shapes. After that, the principal part of the algorithm, the optimization, takes place automatically, during which we allow the \src{} mesh vertices to shift their positions to fit the \tgt{} shape. We optimize an objective function that expresses the Euclidean distance between the \src{} mesh and the \tgt{} geometry, regulated by the marked user correspondence points. The objective function encapsulates complementary terms that concurrently fit the \tgt{} shape and respect the structure of the \src{} mesh. In a nutshell, the process iteratively projects the \src{} mesh onto the \tgt{} surface, and shifts the projected vertices to preserve face angles and local area entropy.

During the optimization phase, our method learns the parameters of a deep neural network that performs the pairwise mapping of \src{} mesh vertex positions to offsets that fit their \tgt{} shape locations. To facilitate the learning process, we introduce a progressive positional encoding layer \cite{hertz2021sape} at the head of the network. Simply put, a progressive encoding layer maps the input vertex positions to a higher dimensional space by feeding them through periodic functions with gradually increasing frequencies.
\ah{During optimization, it progressively reveals higher frequencies of \src{} vertices' positional embedding as an input to the mapping network.} We assert the claims of \cite{tancik2020fourfeat, hertz2021sape} and demonstrate how the optimization benefits both from the stability introduced by spectral bias of the network \cite{rahaman2019spectral} and its non-biased solutions pertaining to high frequency mappings \cite{sitzmann2020implicit, mildenhall2020nerf}.



In Section~\ref{sec:method_ui}, we  describe the preliminary setup of our method.
In Section~\ref{sec:method_loss}, we describe in detail the optimization terms of our \textit{mesh transfer} solution. Finally, in Section~\ref{sec:method_ppe}, we lay out the progressive configuration which allows \ourmethod{} to achieve stable optimization and high quality results.

\subsection{Correspondence Setup}
\label{sec:method_ui}
A key aspect of our method is that it provides the user means to guide the mesh transfer.
First, the user marks a small set of correspondence points $\left\{v_i, u_i\right\}_{i=1}^k$ between the \src{} mesh $\msrc$ and the \tgt{}  shape $\mtgt$.
The correspondence points enable the estimation of an initial global affine transformation from $\msrc$ to $\mtgt$, followed by a biharmonic deformation \cite{jacobson2011bounded} that utilizes the corresponding points on $\mtgt$ as the boundary conditions.
In addition, the user can specify the correspondences as \textit{rigid}, for example in Figure~\ref{fig:bodies}, the movement of the body parts is specified by rigid points. In those cases, we apply another \textit{as-rigid-as-possible} deformation \cite{sorkine2007rigid} using the rigid points as the boundary conditions. See Section~\ref{sec:implementation} for elaborate implementation details.

The optimization phase starts when the user is satisfied with the initial deformation.
The objective of the optimization is to bring the surface of $\msrc$ close to $\mtgt$ while maintaining the structural properties of the original mesh, as described in Section~\ref{sec:method_loss}.

\textbf{Interactive Mode.} \ourmethod{} also includes an interactive mode. Between the optimization epochs, the user may pause, and make additional local modifications by adding or adjusting the correspondence points.


\subsection{Neural Optimization}
\label{sec:method_loss}

\ah{Instead of formulating an \textit{explicit} optimization term between the \src{} mesh and the \tgt{}, we use a neural network parameterization. Specifically, we use fully connected neural networks (FC) with parameters $\theta$. Doing so has been shown to improve the optimization in various works \cite{williams2019deep,Hanocka2020point} as the network serves as an ''internal prior'' in the optimization. }

The mesh deformation is obtained through a direct neural optimization of the parameters $\theta$ of a mapping function $f\left(\msrc \,|\, \theta \right) = \omsrc$ that receives a \src{} mesh, $\msrc$, and outputs an optimized mesh $\omsrc$.

The loss term follows directly from the definition of our problem:
\begin{equation}
\mloss\left(\omsrc \,|\, \mtgt, \ \msrc\right) = \mloss_{d}\left(\omsrc \,|\, \mtgt\right) + \lambda \mloss_{s}\left(\omsrc \,|\, \msrc\right).
\end{equation}
 \ah{On the one hand, we would like our output mesh $\omsrc$ to fit a given target shape $\mtgt$ as closely as possible, }
i.e., minimize the distance $\mloss_{d}\left(\omsrc \,|\, \mtgt\right)$ between $\omsrc$ and $\mtgt$. On the other hand, we wish to preserve the structural quality of the \src{} mesh, which is measured by $\mloss_{s}\left(\omsrc \,|\, \msrc\right)$.

The distance loss is given by 
\begin{equation}
 \mloss_{d}\left(\omsrc \,|\, \mtgt\right) = Ch\left(\omsrc,\ \mtgt\right) + \sum_{i=1}^{k}\|\widehat{v}_{i} - u_{i}\|_{2}^{2} ,
 \label{eq:distance_loss}
\end{equation}
where $Ch(\omsrc, \mtgt)$ is a symmetric Chamfer distance between uniformly sampled points on the optimized mesh and the \tgt{} shape. In addition, we keep the $k$ correspondence points specified by the user close by minimizing the squared distance between them, where $\widehat{v}_{i}$ and $u_{i}$ are pairs of corresponding points on $\omsrc{}$ and $\mtgt$, respectively.

The structural loss  is given by 
\begin{equation}
 \mloss_{s}\left(\omsrc \,|\, \msrc\right) =  \dfrac{1}{N} \sum_{i=1}^{N} \left( \sum_{j \in R(i)} \| \hat{\alpha_j} - \alpha_{j} \|_{2}^{2} + D_{KL}\left(P_i || Q_i\right)\right),
 \label{eq:structural_loss}
\end{equation}
where the summation is over the $N$ vertices of $\msrc$.
The first term represents the distortion of the angles  ${\hat{\alpha_j}}$ on $R(i)$: the 1-ring of $\hat{v}_i$ with respect to the original angles on $\msrc$.
The second term measures the local area Kullback–Leibler divergence where $P_i$ are the fixed areas of the faces around vertex $i$ in $\msrc$, normalized to have sum $1$. $Q_i$ is the equivalent (non fixed) local area distribution in $\omsrc$. \ah{Figure~\ref{fig:ablation_loss} shows the effect of each structural loss term on the final optimization result.}

To prevent numerical issues caused by \textit{skinny} faces, we utilize the quality measure for a triangular face from \cite{liu2020neural}:
\begin{equation}
Q_f = \dfrac{4 \sqrt{3} A_{f}}{\|e_1\|^2 + \|e_2\|^2 + \|e_3\|^2},
\end{equation}
where $A_f$ is the area of the face and  $\|e_i\|$ is the length of its $i$-th edge. When $Q \rightarrow 0$, the face approaches to degenerate zero area. To prevent such cases we penalize by $1-Q_f$ all the faces in $\omsrc{}$ with quality $Q_f < 0.1$.




\subsection{Progressive Positional Encoding}
\label{sec:method_ppe}

It has been shown that the learning bias of deep ReLU networks tends to be towards low frequency functions,  e.g., they have a spectral bias \cite{rahaman2019spectral}. The spectral bias of the network has a positive trait of preventing large deformations during the optimization process. These may be caused by an unstable Chamfer term, which leads to an erroneous local minimum.
Unfortunately, the spectral bias also comes with a price: as the optimization proceeds, it prevents  local \textit{delicate} deformations that bring the surface of the \src{} close to the the \tgt{} shape. That implies that common \net{} with ReLU activations have a hard time mapping a continuous domain to a high frequency image.

To mitigate the deficiencies of \net{} with ReLU activations, previous works by \cite{tancik2020fourfeat, mildenhall2020nerf, sitzmann2020implicit} suggested  frequency based encodings. Specifically in a mesh transfer scenario, \src{} mesh vertex positions are first mapped via positional encodings (PE) before feeding them as input to the \net. However, in the case of mesh transfer, static frequency encoding schemes introduces a new problem to the architecture. Encoding functionals of high frequencies may overfit too quickly, causing the optimization to converge to suboptimal solutions. See for example, Figure~\ref{fig:model_compare}, for the distortion caused by the positional encoding neural optimization.

Instead, \ourmethod{} leverages the progressive positional encoding layer of \cite{hertz2021sape}.
Progressive positional encoding 
operates under the assumption that earlier iterations of the optimization benefit from the spectral bias, which is achieved by low frequency encodings. As the optimization converges, higher frequencies encodings are used to fit the delicate features of the shape.
\op{In the context of mesh transfer, this formulation enforces an inductive bias which ensures the optimization is both stable and accurate. In our experiments we adopt a lightweight version of \cite{hertz2021sape}: we leverage progressive positional encodings, and trace their progression in a global manner, e.g: all vertices are exposed to higher frequencies at the same time-step.}
\renewcommand{\labelitemi}{$\textendash}

\begin{figure*}
\begin{overpic}[width=\textwidth,tics=10]{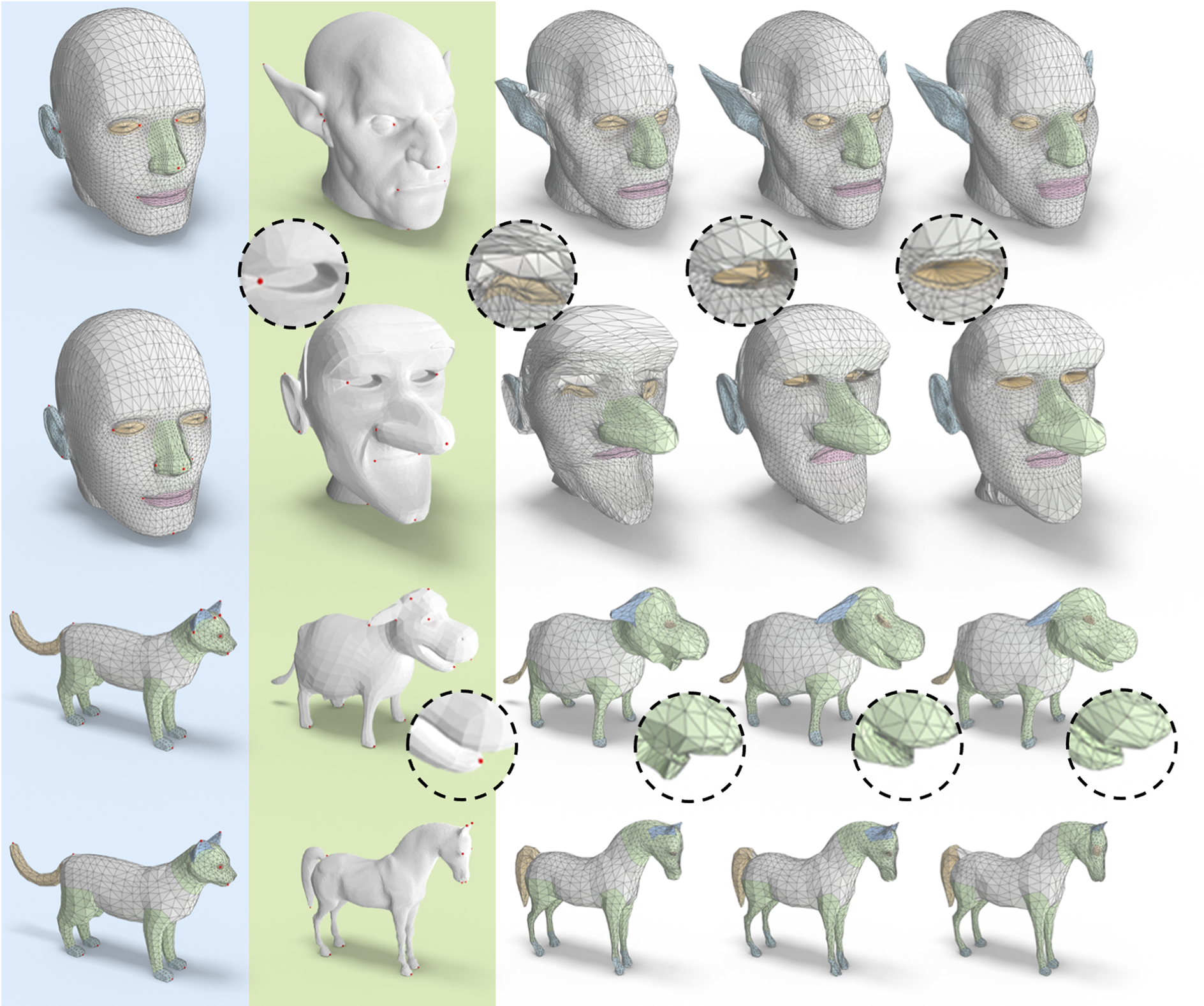}
\put(8,1){\small{Source}}

\put(29,1){\small{Target}}

\put(43,1){\small{Elastic Correspondence}}

\put(64,1){\small{Inter-Surface}}

\put(82.5,1){\small{\ourmethod{}}}
\end{overpic}

 \caption{Comparison to direct surface mapping methods. Sparse correspondence points manually marked by users to guide algorithms are indicated by red dots. Face colors indicate semantic areas, showing how the mesh is stretched or contracted to fit the target shape. \ourmethod{} is able to handle transfer of fine mesh structure in highly detailed areas.}
 
 \label{fig:compare_mapping}
\end{figure*}
\begin{figure*}
\begin{overpic}[width=\textwidth,tics=10]{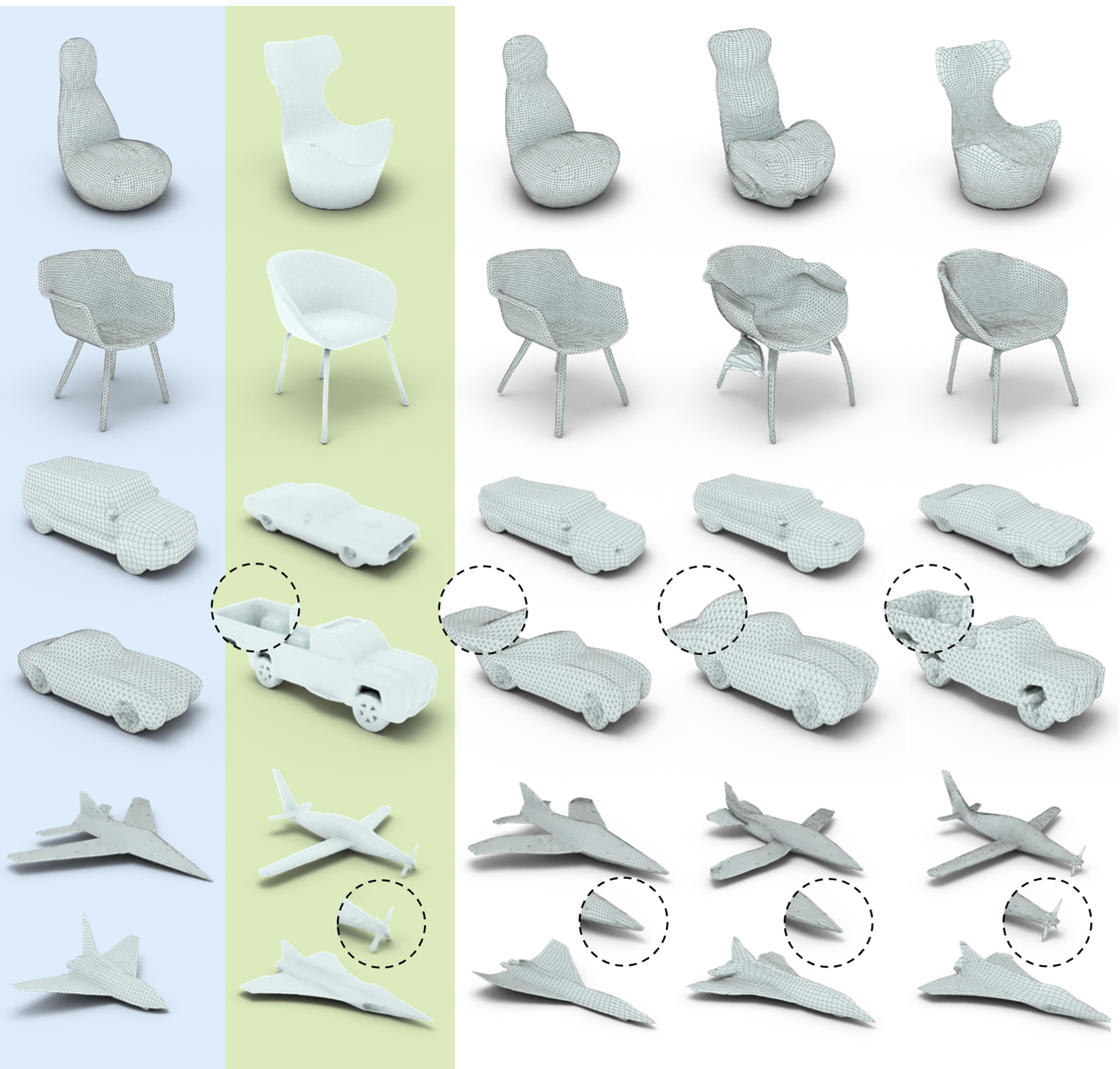}
\put(8,1){\small{Source}}

\put(28,1){\small{Target}}

\put(47, 1){\small{Neural Cages}}

\put(67,1){\small{ShapeFlow}}

\put(87.5,1){\small{\ours{}}}
\end{overpic}

\caption{Comparison to deep deformation methods. The usage of progressive positional encoding allows \ours{} to deform the source mesh and remain faithful to delicate details of the target shape. }

\label{fig:compare_deform}
\end{figure*}









\begin{figure*}
\begin{overpic}[width=\textwidth,tics=10]{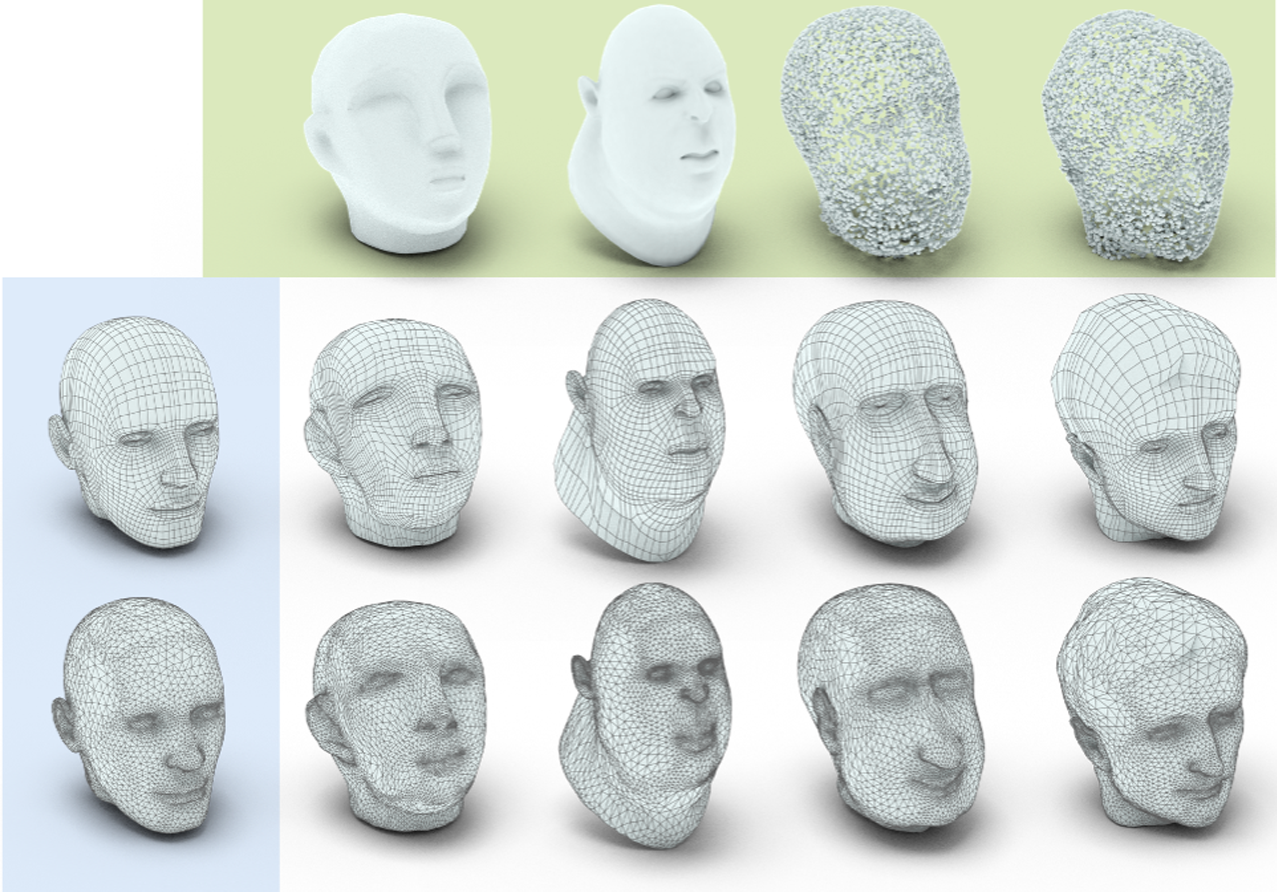}

\put(16.5,50){
\small{Target}
}

\put(9.5,1){\small{Source}}
\end{overpic}

     
     \caption{Applying \ourmethod{} to transfer the structure of the mesh faces on left to the \tgt{} meshes or point clouds on top. Our method supports direct mesh transfer to non-manifold shapes.}
     \label{fig:faces}
\end{figure*}
\begin{figure}
\includegraphics[width=\columnwidth]{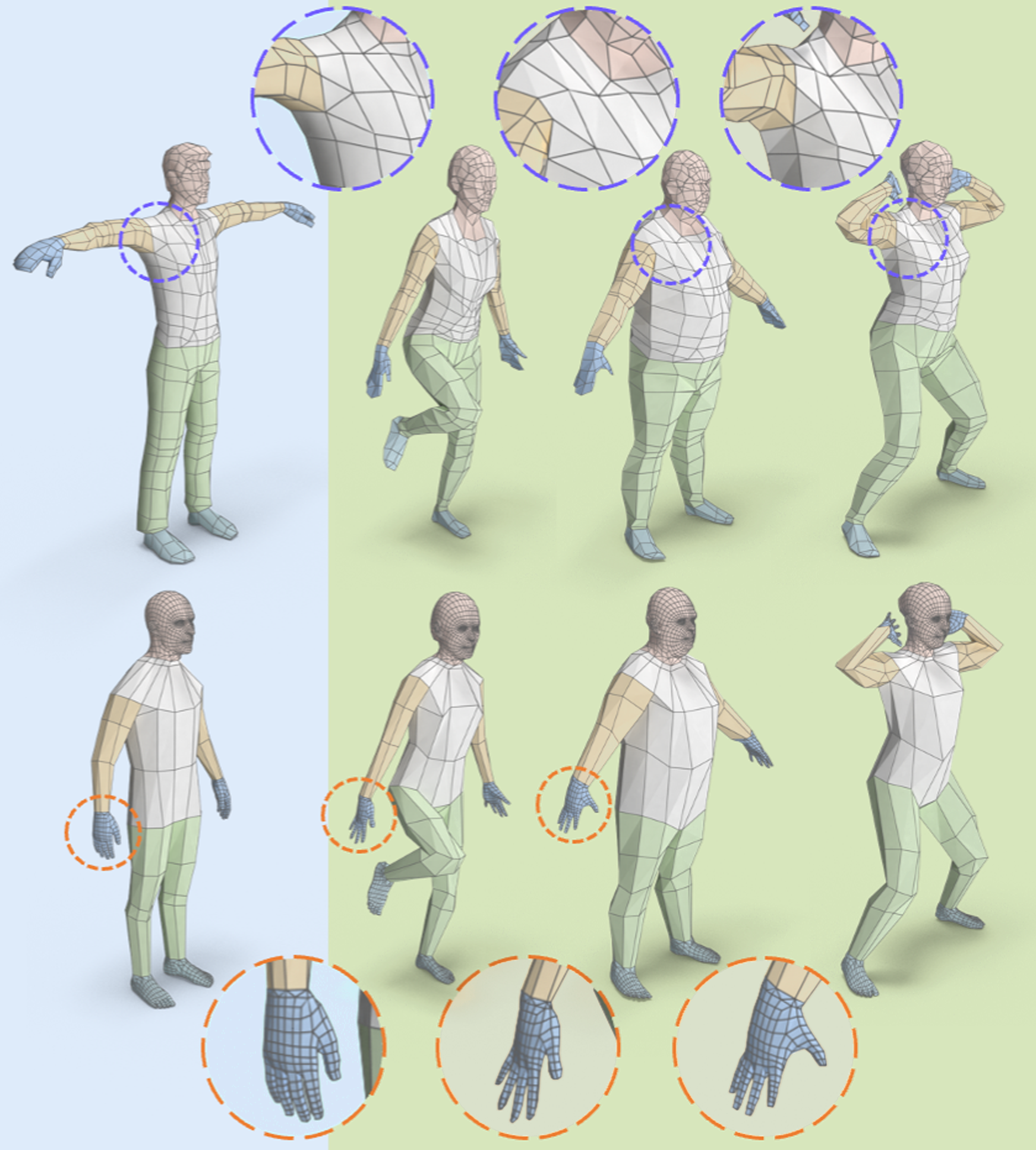}


        \vspace{-8pt}
     \caption{Mapping meshings of varying density, highlighted in different colors. \ourmethod{} is able to fit the target shape while retaining the original mesh structure with minimal distortion (zoom in for details). }
     \label{fig:bodies}
     \vspace{-10pt}
\end{figure}

\section{Evaluation}
\label{section:experiments}
Unless specified otherwise, we use the same optimization configuration detailed in Section~\ref{sec:implementation} for all described experiments. 

\subsection{Implementation Details}
\label{sec:implementation}

\textbf{Correspondence Setup.}
For most results shown in this paper, our algorithm requires users to mark 8-15 pairs of correspondence points on average. This number may vary with the complexity of the models. As a good measure, up to 10 correspondence points is the amount used to generate all face figures within this paper and up to 20 for the human bodies in Figure \ref{fig:bodies}.
For processing the influence area of correspondence points, we use the implementation of \cite{jacobson2011bounded} and \cite{sorkine2007rigid} from \emph{libigl} \cite{libigl}.

\textbf{Network Architecture and Optimization.}
Our architecture consists of an \net{} with $4$ hidden layers, of size $256$, where the first layer is a progressive positional encoding \cite{hertz2021sape} layer divided into $6$ blocks.
Our optimization runs for $1500$ iterations, alternating between backpropagations on $\mloss_{d}\left(\omsrc | \mtgt\right)$ and $\lambda\mloss_{s}\left(\omsrc | \msrc\right)$, where the weight $\lambda$ is set to $1$ for the first $1000$ iterations and later lowered to $0.2$ for the rest of the optimization.


%
\textbf{Latency.} On a machine equipped with Nvidia GTX 1080, the optimization takes up to 45 seconds to converge for $1500$ iterations.


\subsection{Evaluation Metrics}
\label{sec:exp_quality}
\begin{figure*}

\begin{overpic}[width=\textwidth,tics=10, trim=0 0 0 0,clip]{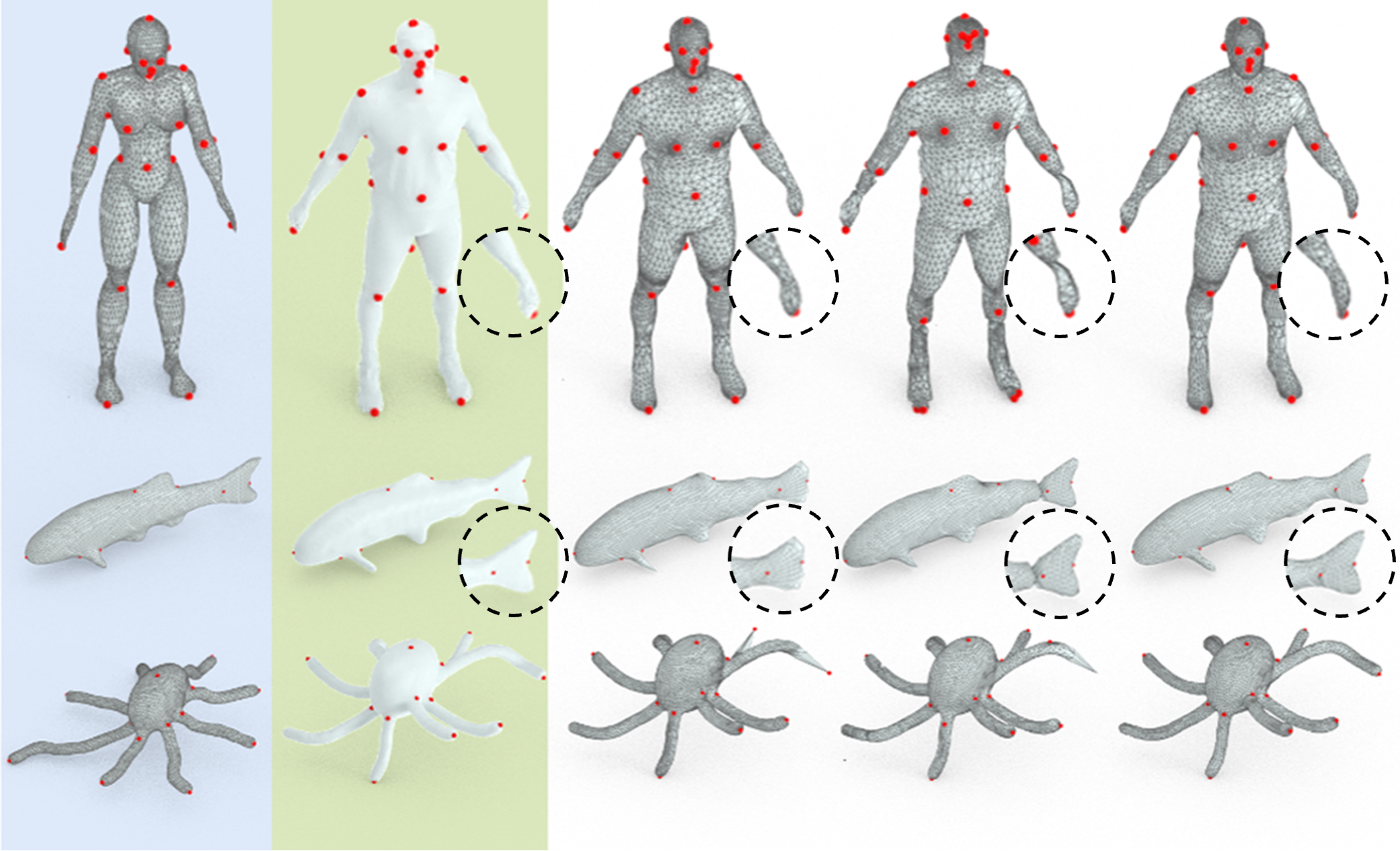}

\put(8,1){{Source}}

\put(26.5,1){{Target}}

\put(47,1){{HOT}}

\put(66,1){{RHM}}

\put(83.5,1){{\ourmethod{}}}

\end{overpic}

 \caption{Comparison to additional surface mapping methods on the SHREC-BIM benchmark. In this comparison, correspondence points are provided with the data and not by manual user interactions. Red dots indicate correspondence points used by each method.}

 \label{fig:compare_shrec}
\end{figure*}
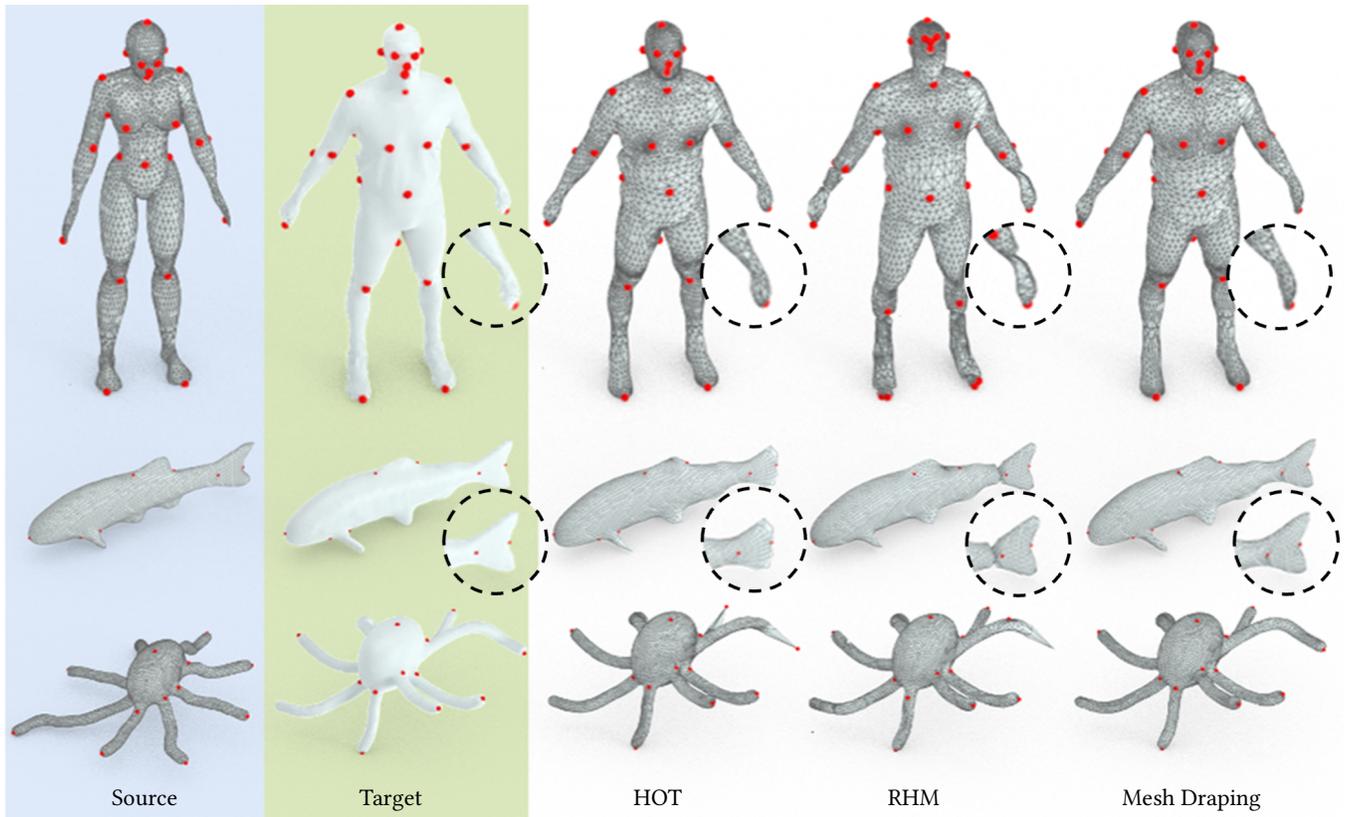

Our quantitative and qualitative evaluations highlight the importance of jointly optimizing a source mesh distortion metric, as well as the alignment with the target shape. We measure the quality of distortion of $\omsrc{}$ with respect to the \src{} mesh $\msrc{}$ by discrete Dirichlet energy \cite{pinkall1993computing}. The alignment integrity of the optimized mesh $\omsrc{}$ over the \tgt{} shape  $\mtgt{}$ is evaluated using Chamfer and Hausdorff distances. It should be noted that the distortion and alignment metrics are complementary to one another, i.e, optimizing one but not the other provides inferior results as either the source structure is distorted, or the result does not perfectly align with the target shape. Readers may confirm the last statement by cross-comparing the visualizations of Figure~\ref{fig:compare_mapping}, Figure~\ref{fig:compare_deform} and Figure~\ref{fig:compare_shrec} with the quantitative evaluation in Table \ref{tab:transfer_metrics}.

The aforementioned metrics are staple, and are commonly used on an individual basis. However, when measured concurrently for mesh transfer quality assessment, one has to account for their scale differences, their specific ranges, and how to effectively quantify them into one, comparable measurement. To that end, we propose a novel evaluation metric, \quality{}, which combines source distortion and target alignment measurements to a single score. 

We define \quality{} using the following template function:
\vspace{-5pt}
\begin{equation}
 \mathbf{Q}_{\tau}(\omsrc \,|\, \msrc,\ \mtgt) = 1 - \exp\left(\frac{-\tau}{\| \mfunc{}_{d}(\msrc, \omsrc) + \mfunc{}_{a}(\mtgt, \omsrc) \|_{2} }\right).
\end{equation}
Here, the notations follow Section~\ref{sec:method_loss}, where $\msrc$, $\mtgt$ and $\omsrc$ represent the source mesh, the target shape and the transferred mesh, respectively.

$\mfunc{}_{d}$ and $\mfunc{}_{a}$ are respectively the source distortion and target alignment measure functions, which are chosen in this paper as the Dirichlet energy and Hausdorff distance defined as 
\begin{equation}
 \mfunc{}_{d}(\msrc, \omsrc) = \dfrac{1}{2} \left(\sum_{f \in \msrc} | d\phi_{\msrc\omsrc}(f) |_{2}^{2} \, a_{f} \right) - 1,
\end{equation}
%
%
%

\begin{equation}
 \mfunc{}_{a}(\mtgt, \omsrc) = w_a \max \Big\{ \adjustlimits \sup_{\vphantom{\omsrc}x \in \mtgt} \inf_{y \in \omsrc} \mathrm{d}(x,y), \adjustlimits \sup_{x \in \omsrc}  \inf_{y \in \mtgt} \mathrm{d}(x,y) \Big\} ,
\end{equation}
where for the discrete Dirichlet energy, we follow the definition of \cite{ezuz2019reversible} and assume $d\phi_{\msrc \omsrc}(f) \in \mathbb{R}^{2 \times 2}$ is the linear transformation
that maps face $f \in \msrc $ to its image in $\omsrc$, and $a_f$ is the face area. To map the Dirichlet energy to the range of $[0,\infty]$, we also subtract a single unit. To align the magnitude of the Hausdorff distance, we scale it by a constant $w_{a}=1e^2$, which we set empirically.

The calibration hyper-parameter $\tau$ is used to tone down the scale of the metric to better distinguish how different methods compare to each other. Throughout this paper, we use $\mathbf{Q}_\text{transfer} = \mathbf{Q}_{\tau}$ with $\tau=5$.
%
%
In practice, prior to metrics computation, all 3D models are normalized to the unit cube to ensure scale invariance.  

To conclude, \quality{} exhibits the following attributes:
\renewcommand{\theenumi}{\roman{enumi}}%
\begin{enumerate}
    \item \quality{}$\in[0, 1]$, where a higher score correlates with better quality.
    \item A perfect score of 1.0 is obtainable when $\mfunc{}_{d}=1$ (no distortion occurs) and $\mfunc{}_{a}=0$ (optimized shape perfectly aligns with target).
    \item When $\mfunc{}_{d} \to \infty$ or $\mfunc{}_{a} \to \infty$, then \quality{}$=0$.
    \item A lower $\tau$ yields a ``harder'' evaluation metric, which penalizes to a greater extent both misalignment and high distortion.
\end{enumerate}

To tie up the discussion of metrics, we acknowledge that one of the terms in our evaluation metrics, the Chamfer distance, is directly used as one of the optimization terms. We emphasize this choice does not impact the integrity of the evaluation due to the optimization objective encompassing additional terms, and our evaluation consisting of other metrics as well.


\begin{table*}
\centering
\ra{1.0}
\setlength{\tabcolsep}{2.99pt}
\begin{tabular}{@{}llcccccccccc@{}}
\toprule

\emph{Evaluation} && \multicolumn{2}{c} {\textbf{Alignment}} &  \textbf{Distortion} & \textbf{Joint} \\

  \emph{Metric} && Chamf. $(1e^{-4})$ \scriptsize{($\downarrow$)} & Hausd. $(1e^{-2})$ \scriptsize{($\downarrow$)} & Dirichlet \scriptsize{($\downarrow$)} & \quality{} \scriptsize{($\uparrow$)} \\ \midrule
Elastic Corresp. && 1.97 & 4.23 & \textbf{1.52} & 0.690 \\
  Inter-Surface && 1.39 & 3.14 & 2.29 & 0.770  \\
  Ours && \textbf{1.18} & \textbf{1.91} & 1.84 & \textbf{0.900} \\
 \midrule
 HOT && 13.78 & 4.57 & 12.76 & 0.327 \\
 RHM && 8.78  & 3.49 & 6.54 & 0.534
  \\
  Ours && \textbf{8.34} & \textbf{2.22} & \textbf{3.29} & \textbf{0.791} \\
 \midrule
  Neural Cages && 38.68 & 11.16 & \textbf{1.1} & 0.361 \\
  ShapeFlow && 12.81 & 8.77 & 4.8 & 0.407 \\
  Ours && \textbf{1.84} & \textbf{1.74} & 1.6 & \textbf{0.933}   \\
\bottomrule
\end{tabular}
\vspace{5pt}
\caption{
Quantitative comparison of our approach to other surface mapping and deformation methods, on the task of mesh transferring. We measure the mean Chamfer distance, Hausdorff distance, Dirichlet energy, and our proposed \quality{} metric over the shapes shown in Figures~\ref{fig:compare_mapping}, \ref{fig:compare_deform} and \ref{fig:compare_shrec}.
\label{tab:transfer_metrics}
}
\end{table*}

\subsection{Comparisons}
\label{section:comprison}
We compare our method to other methods that we found as the most relevant to the task of mesh transfer. The quantitative results are summarized in Table~\ref{tab:transfer_metrics}.

\textbf{SHREC-BIM benchmark.}
We evaluate our method on the BIM benchmark \cite{kim2011blended} which includes 200 pairs of meshes from SHREC dataset \cite{giorgi2007shape}. Each pair is supplied with with corresponding points (between 2-36 points).
We compare our method with two parameterization based
methods: Hyperbolic Orbifold Tutte Embeddings \cite{aigerman2016hyperbolic} (HOT) and Reversible Harmonic Maps between Discrete Surfaces \cite{ezuz2019reversible} (RHM). Both methods receive as input the corresponding points and output a parametrization between the meshes that maps the vertices of one mesh to barycentric coordinates on the another mesh. As can been seen in Figure~\ref{fig:compare_shrec}, the parametrization methods may struggle to keep on the triangulation of the source mesh. \ah{For example, see artifacts RHM outputs such as flipping faces on the fish tail and humanoid hand.}

\textbf{Additional mapping methods.}
We compare our method to additional two surface mapping methods on our custom test set. The first is Elastic Correspondence (\elc)  \cite{ezuz2019elastic} using the initialization scheme of Aigerman et al. \cite{aigerman2016hyperbolic}, and the second is Inter-Surface Maps due to (\ism) \cite{schmidt2020inter}. Both methods are based on parameterization. Consequentially, they are only applicable for pairs of meshes, and also assume the existence of some bijectivity between the surfaces.

The comparisons are shown in Figure~\ref{fig:compare_mapping} where the input to the three methods are pairs of \src{} and \tgt{} meshes with their marked corresponding points. As evident by careful observation of the corresponding segmented parts marked on the \src{} and the optimized mesh, \elc{} causes semantic distortion, for example, on the top face ears or the cow mouth.
Both \ism{} and our method preserve the semantic correspondence between the meshes.
This observations are also reflected in the quantitative results. Due to some distortions in the mapping of \ism, our method achieves \textit{slightly better} results. 

It is important to mention, that both methods, \ism{} and \elc{}, satisfy a complete bijective map between the input pair of meshes. However, an error evolves in places where the new discrete tessellation of the \src{} does not properly cover the \tgt{} when projecting the map onto the \tgt{} mesh. We suspect that these kind of errors appear mostly at delicate places, i.e, at the eyes of the topmost face and the cow mouth (Figure~\ref{fig:compare_mapping}). In comparison, our direct optimization avoids such types of distortion.
 

\textbf{Comparison to deformation methods.}
We also compare our method to two deformation methods. Both methods are based on a pre-training step over some specific dataset. Those methods can be categorized by the priors they inject to the deformation allowed between the \src{} to the \tgt{} shapes. The first method, Neural Cages (\ncage) \cite{yifan2020neural}, allows only coarse deformations by displacing the control points of a \textit{cage} that wraps the \src{} mesh. The second method, ShapeFlow (\sflow) \cite{jiang2020shapeflow}, enables a more flexible deformation by allowing displacement for all vertices of the \src{} mesh. Their deformation is regularized through the loss function that includes a volume conservation term and a penalization term for a non isometric deformation. 

The results are shown in Figure~\ref{fig:compare_deform}, where we trained and tested each method on selected classes from the ShapeNet dataset \cite{chang2015shapenet}.
We apply our method without providing correspondence points, , \op{e.g: we set $k=0$ in Eq.~\ref{eq:distance_loss}. In other words, we assume only a global orientation} alignment between the \src{} and the \tgt{} meshes, as both other methods assume.
Both qualitative and quantitative comparisons clearly highlight the difference between the methods. Our proposed progressive optimization better aligns the \tgt{} mesh with the target shape, doing so with few distortions. By utilizing only global deformations, \ncage{} minimizes the distortion term but the resulted mesh is unable to satisfactorily obtain the form of the \tgt{} mesh. Finally, \sflow{} brings the \src{} mesh closer to the form of the \tgt{} shape, but the resulting meshes usually suffer from undesired distortions.

\textbf{Polygon soups and point clouds targets.}
Finally, we highlight the strengths of our progressive optimization by presenting visual examples of mesh-transfer in non-trivial scenarios. In Figure~\ref{fig:faces}, we demonstrate a transfer of face meshes to point clouds, a task which surface mapping methods relying on bijective mappings cannot directly perform. The quality of our results are on par with the mesh-to-mesh case. This setting pertains to a real-life application, where experts may reuse predefined meshes on recently scanned objects.

\textbf{Varying density meshes.} In Figure~\ref{fig:bodies}, we illustrate the example of transferring meshes of varying density. For clarity, different areas of interest are segmented by color. Though we suggest a global approach, it is able to transfer both sparsely and densely detailed polygonal areas with minimal distortion, maintaining the original intention of the expert artist. This example alludes to the case where specific parts of the mesh are highly detailed for a particular purpose, such as animation.











\begin{figure}
       \begin{overpic}[width=\columnwidth,tics=10]{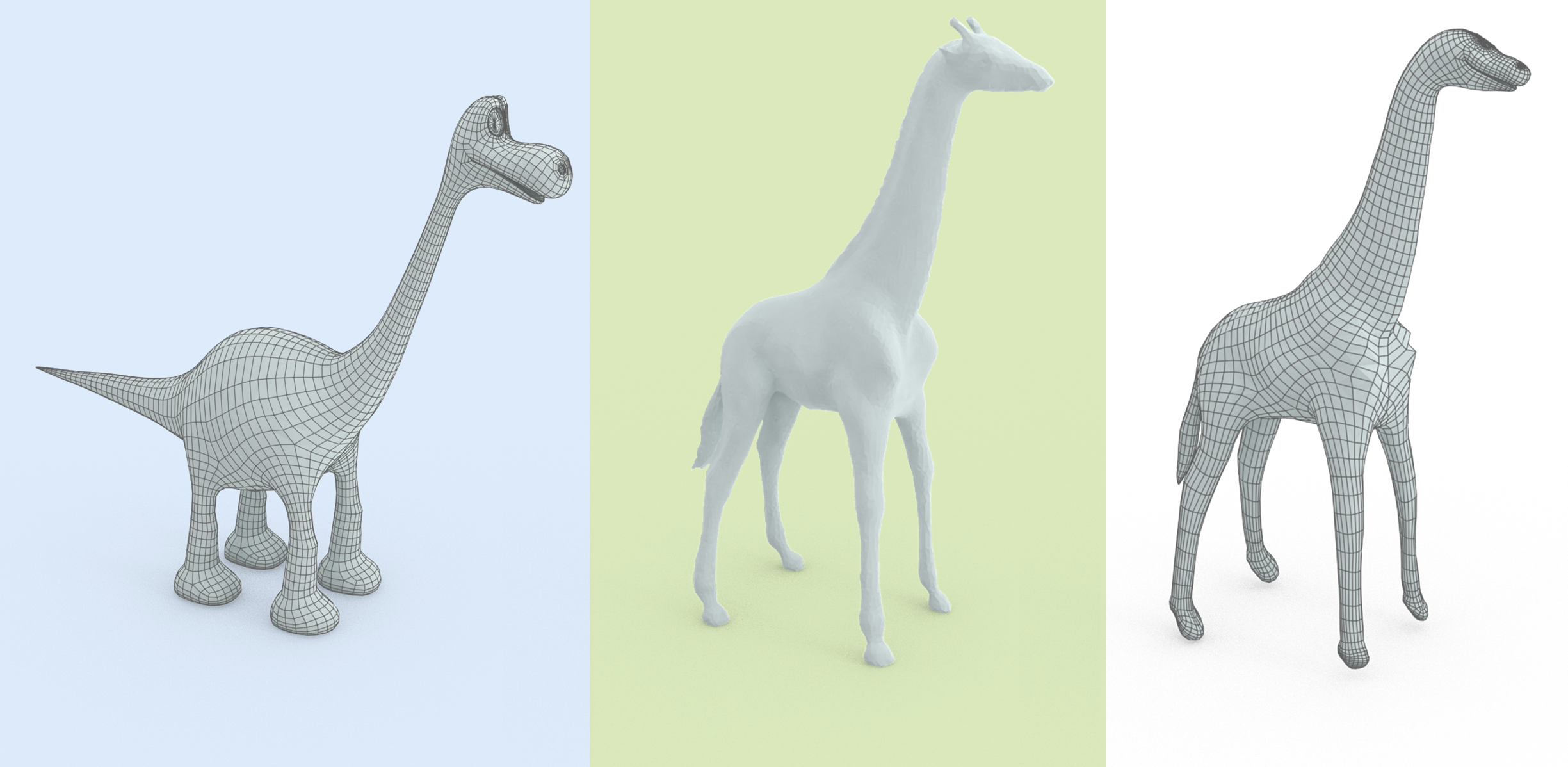}

    \put(12.5,1.5){Source}
    \put(49,1.5){{Target}}
    
    \put(73, 1.5){\ourmethod{}}
    \end{overpic}

    \caption{\op{Limitation example: mesh transfer from the dinosaur mesh to the giraffe shape yields sub-optimal results unfaithful to the original target shape. Due to large shape differences, the tessellation of the dinosaur inherently \textit{struggles} to accurately fit the giraffe ears or the delicate structure of its feet.}}
    \label{fig:limitation}
    
\end{figure}

\textbf{Limitations.}
%
While \ourmethod{} exhibits very good results in many cases, there are still some things to improve in it. 
First and foremost, \ourmethod{} is intended for pairs of shapes of similar parts. \ah{For example - mesh transfer will succeed between different four legged animals (Figure~\ref{fig:teaser}), but will under-perform when transferring meshing from humans to animals, or from a dinosaur to a giraffe as shown in Figure~\ref{fig:limitation}.
Here, the horns shape of the giraffe requires special meshing, which does not exist on the dino head model.}
This phenomenon is also demonstrated in Figure~\ref{fig:compare_deform}, row 4: the target truck shape contains a deep depression right above the right-front wheel, whereas the source shape is smooth. As a result, part of the resulting wheel mesh appears to be deformed.
In addition, on extreme cases of sharp edges (so called "spikes"), the method may struggle to achieve perfect alignment between the optimized mesh and the target shape. To maintain high quality results on very detailed meshes, users may be required to define additional correspondence points. Since we do not use a pre-defined train set, in such cases user intervention is necessary.

By design, our method does not alter local connectivity at all. However, where needed, one may perform local subdivision or remeshing as a postprocess.
Finally, the latency measures in this paper, are presented for an unoptimized implementation. Subsequent efforts may reduce the runtime cost (detailed in Section~\ref{sec:implementation}) by further optimizing the \ours{} logic.
\section{Conclusion and Future Work}

We presented a parameterization-free approach for transferring mesh structure to comparable shapes. The proposed method uses neural optimization with progressive frequency positional encodings, which contribute to both stable optimization and high quality fitting to fine details.  We demonstrated the applicability of our method on a range of examples, including the re-use of triangular and quad meshes to target meshes and point clouds.

Future work may leverage an unpaired training set of shapes to obtain priors on common shapes for faster and more robust optimization.
Another direction for future work may improve the computation time of our optimization, or its flexibility by allowing users to guide mappings between different topologies, or jointly deform a shape from several sources.

\bibliographystyle{ACM-Reference-Format}
\bibliography{bibs}

\end{document}